\documentclass[conference]{IEEEtran}
\IEEEoverridecommandlockouts
\usepackage{graphics}
\usepackage{graphicx}
\graphicspath{{Graphs/}}

\begin{document}

\title{Outlier Elimination for \\Robust Ellipse and Ellipsoid Fitting\thanks{This research was supported in part by the Office of Naval Research and the US Army Research Office under Grants N00014-09-1-0342, W911NF-07-1-0185 and N00014-07-1-0555.}}

\author{\IEEEauthorblockN{Jieqi Yu}
\IEEEauthorblockA{Dept. of Electrical Eng.\\Princeton University\\
Princeton, NJ, 08544\\
jieqiyu@princeton.edu}
\and
\IEEEauthorblockN{Haipeng Zheng}
\IEEEauthorblockA{Dept. of Electrical Eng.\\Princeton University\\
Princeton, NJ, 08544\\
haipengz@princeton.edu}
\and
\IEEEauthorblockN{Sanjeev R. Kulkarni}
\IEEEauthorblockA{Dept. of Electrical Eng.\\Princeton University\\
Princeton, NJ, 08544\\
kulkarni@princeton.edu}
\and
\IEEEauthorblockN{H. Vincent Poor}
\IEEEauthorblockA{Dept. of Electrical Eng.\\Princeton University\\
Princeton, NJ, 08544\\
poor@princeton.edu}}

\maketitle

\begin{abstract}
In this paper, an outlier elimination algorithm for ellipse/ellipsoid fitting is proposed. This two-stage algorithm employs a proximity-based outlier detection algorithm (using the graph Laplacian), followed by a model-based outlier detection algorithm similar to random sample consensus (RANSAC). These two stages compensate for each other so that outliers of various types can be eliminated with reasonable computation. The outlier elimination algorithm considerably improves the robustness of ellipse/ellipsoid fitting as demonstrated by simulations.
\end{abstract}


\IEEEpeerreviewmaketitle

\section{Introduction}

    Although there are many off-the-shelf algorithms for ellipse and ellipsoid fitting (e.g., \cite{Fitzgibbon1999} \cite{Ahn2001} \cite{Chojnacki2000}), there are some limitations when applying these algorithms to data with many highly noisy observations (outliers). Therefore, before fitting the data, it is desirable to eliminate the outliers first. To this end, we classify a group of unlabeled data into two classes. One of them consists of data that can be fit well by an ellipse (ellipsoid), and the other consists of data that can be classified as outliers. In other words, it is a two-class classification problem with prior knowledge on one of the classes. 

    As a well-developed area, outlier detection has been studied from several perspectives, such as statistics, neural networks and machine learning \cite{Hodge2004}. However, due to the special properties of the ellipse fitting problem, it is rather difficult to solve using merely one type of outlier detection algorithm. 
    
    One difficulty is that the two classes, i.e. outliers and inliers, usually intertwine in a complex manner, as shown in Fig. \ref{ClassAB}. Some of the outliers are far from the ellipse (like point A), yet others may fall inside the ellipse (like point B). Many outlier detection algorithms based on proximity (e.g., $k$-NN \cite{Ramaswamy2000}) may not be effective for detecting the B-type outliers.
    
    On the other hand, with a high percentage of outliers, if we resort to model-based outlier detection algorithms like random sample consensus (RANSAC) \cite{Fischler1981}, which can better handle the B-type outliers, the large number of parameters of the model looms as a serious problem. On defining $w$ as the portion of the inliers, $n$ as the minimum number of data points needed to fit a model, and $p$ as the probability of successfully finding the correct model after running RANSAC $k$ times, we have the following relationship: $p = 1 - (1-w^n)^k$. When $w$ is small (the percentage of outliers is high), we need $k$ to be large to have a sufficiently high $p$. Unfortunately, the increase in $k$ with respect to $n$ is exponential. If we assume $w=0.5$, to guarantee $p=0.99$, we need $k=16$ for a simple two-dimensional straight line model; yet for ellipse fitting, where $n=5$, we need $k=146$. For an ellipsoid ($n=9$), we need $k = 2356$. Moreover, when $n$ is large, the fitting algorithm itself becomes rather expensive to run even once.
	
	To counter these problems, in this paper we develop a two-stage hybrid outlier detection algorithm for ellipse/ellipsoid fitting. First, we employ an outlier detection algorithm based on algebraic graph theory \cite{Mohar1991}, using proximity information of the data points. Second, after reducing the portion of the outliers among the remaining data set, we employ a robust, model-based outlier detection algorithm to efficiently refine the results. In the first stage, a large portion of the most distant, isolated outliers, which can be a great hazard to model-based outlier detection methods, are detected. In the second stage, the subtler, closer outliers that go against the ellipse model are detected, and some of the misclassified inliers are recovered. These two stages compensate for each other to form a more efficient and accurate outlier detection algorithm.
	
	The structure of the paper is as follows: In Section II, the model is specified and the three major steps, constructing the graph Laplacian and choosing the parameters, detecting outliers in the eigenvectors, and refining the final results using RANSAC, are described. In Section III, simulation results of outlier elimination for both ellipse and ellipsoid fitting are described. Section IV concludes our work. 

\section{Model and Method}
\subsection{Model: Basic Assumptions}

First, we need to specify the model by stipulating a few assumptions on the data. In our description, we restrict to the case of ellipse fitting, since it is easy to generalize to ellipsoid fitting. Assume that we have a total number of $K$ points $\{(x_i,y_i)\}_{i=1}^K$ with $N$ points from an ellipse with small amounts of noise (inliers) and $M$ points randomly scattered in the plane (outliers). We do not assume a concrete statistical model for the outliers (which is usually hard to do in practice), yet we assume the following:
	\begin{enumerate}
	\item The average distances between inliers are smaller than those between inliers and outliers;
	\item The inliers are the majority ($\geq50\%$) of the data set;
	\item The inliers arise from an ellipse and their noise level is low enough so that it is possible to fit a good model based on a small portion of these inliers (ideally 5 points for the ellipse case). 
	\end{enumerate}
		\begin{figure}[h] 
        \centering{
        \includegraphics[scale = 0.25]{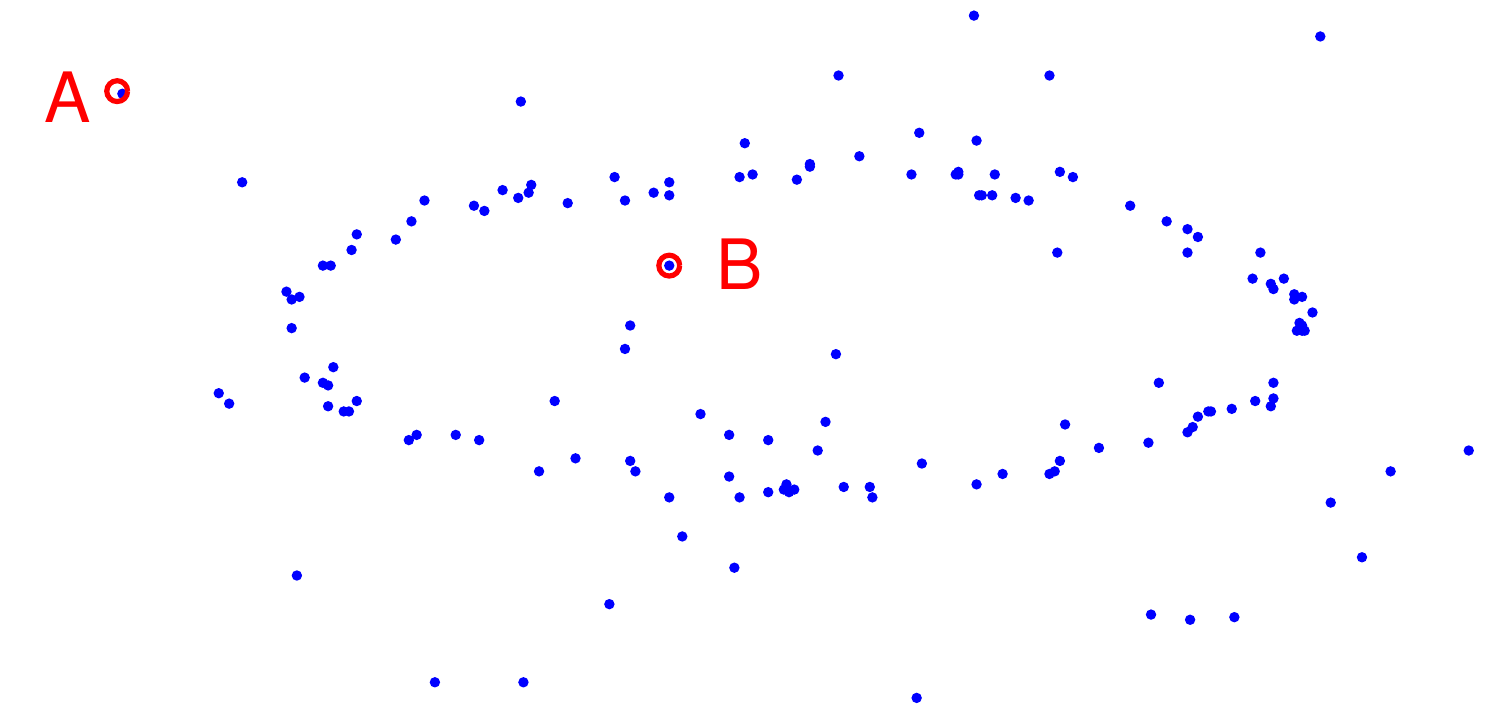}
        \caption{Two different types of outliers. Proximity-based methods are good at detecting outliers similar to point A, while model-based methods can better detect outliers similar to point B.}
        \label{ClassAB}}        
    \end{figure}
	With the first two assumptions, it is possible, by properly selecting the radius of connection, to construct a graph based on proximity in which the inliers are strongly connected to each other, while being weakly connected to the outliers. Then, based on some basic concepts of algebraic graph theory, we can convert the outlier detection problem in a two-dimensional space into a simple one-dimensional real-valued outlier detection problem among the elements of eigenvectors. Moreover, with the third assumption, we can apply a model-based algorithm to refine the detection results of the proximity-based algorithm. 

\subsection{Constructing Graph Laplacian}
Generally speaking, proximity-based outlier detection algorithms depend on an adjacency graph constructed by connecting adjacent points. According to algebraic graph theory, the connectivity information is reflected in the graph Laplacian and its corresponding eivenvalues and eigenvectors.

In our algorithm, instead of using an unweighted graph with a hard connection rule, the graph Laplacian is constructed using heat kernels in the following manner:
	\begin{enumerate}
	\item Form an adjacency matrix $Q$ with $Q_{ij}$ being the Euclidean distance between the points $(x_i,y_i)$ and $(x_j,y_j)$;
	\item Form a fully connected adjacency graph with edge weight $W_{ij}=e^{-Q_{ij}^2/t}$ (heat kernel); 
	\item Construct the graph Laplacian matrix: $L=D-W$, where $D$ is a diagonal matrix with its diagonal entries as column sums (or row sums) of $W$;
	\item Compute eigenvalues and eigenvectors for the generalized eigenvector problem, $L\mathbf{f}=\lambda D\mathbf{f}$.
	\end{enumerate}
Then, the set $\{\lambda    |   \lambda \approx 0, \lambda \in\{\lambda_i\}_{i=1}^K\}$ and its corresponding eigenvectors form the cornerstone of our proximity-based outlier detection algorithm.

\begin{figure}[!hbtp] 
        \centering{
        \includegraphics[scale = 0.4]{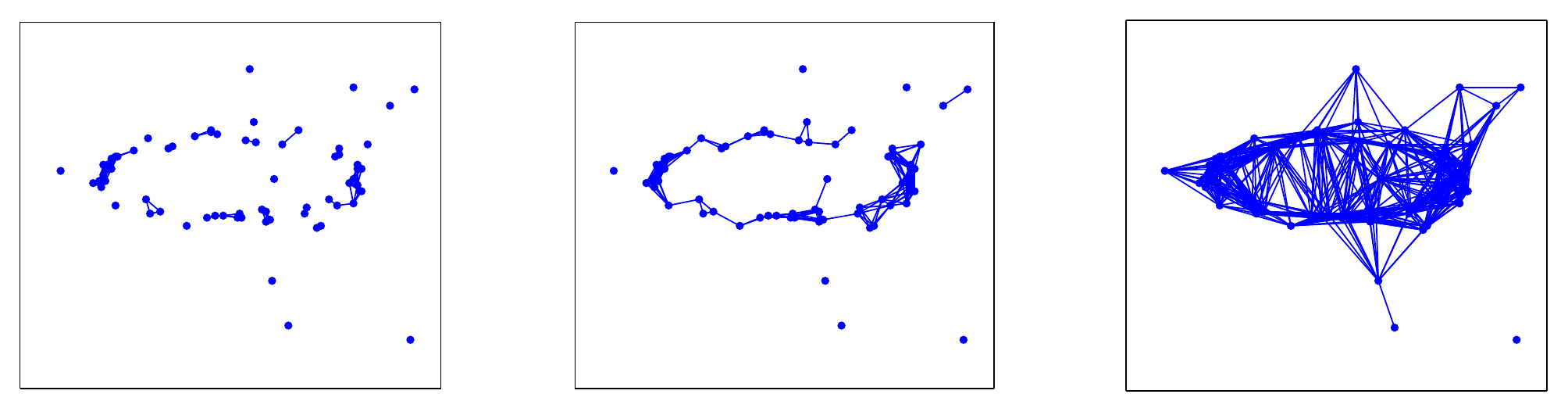}
        \caption{Different choices of connection radius lead to different results. The figure on the left depicts the situation where the radius is so small that even the inliers do not connect sufficiently with each other. The figure on the right illustrates the case where the radius is so large that almost all the data points are connected with each other, and thus makes it impossible to distinguish inliers from outliers. The figure in the middle shows that with a properly selected radius of connection, the inliers form a large connected component, leaving most of the outliers unconnected.}
        \label{Connection}}      
    \end{figure}

The success of the proximity-based outlier detection highly depends on the choice of the ``radius of connection" (i.e., $\sqrt{t}$ in the heat kernel). Thus, the value of $t$ is of essential importance. Different choices of this radius can result in graphs that differ drastically from each other, as shown in Fig. \ref{Connection}. For large $t$, the vertices tend to be strongly connected and hence we might miss many outliers, while for small $t$, the graph tends to be separated into many weakly connected blocks, and we are more likely to misclassify many inliers as outliers.

The choice of $t$ is a rather delicate problem. 
Here we present an intuitive empirical rule to determine a good value of $t$. Sort all the elements of $Q$ in ascending order and denote the resulting vector as $\mathbf{q}$. Then take the $pK$-th element of $\mathbf{q}$ as $\sqrt{t}$. Intuitively, most of the small elements of $Q$ are composed of the distances between inliers, according to our assumptions. By taking the $pK$-th element of $\mathbf{q}$ as $\sqrt{t}$, we approximately guarantee a strongly connected graph among the inliers of average degree of at least $p-1$, if we assume $e^{-1}$ is the threshold for a ``connection". Empirically, $p=4$ works well.

\subsection{Selecting Outliers in Eigenvectors}

Given an appropriate $t$, the inliers form a strongly connected component, leaving most of the outliers loosely connected to it and each other. As a result, with a proper interchange of rows and columns, the adjacency matrix $Q$, and thus the Laplacian matrix $L$, is close to a block diagonal matrix. Each block corresponds to a strongly connected component, with the largest one corresponding to the inliers.

    For each block, there is an eigenvalue approximately equal to zero, and the corresponding eigenvector is almost a binary vector composed of 0's and 1's, with only the elements aligned with the block being 1's. Here, we assume that the $L^\infty$ norms of all the eigenvectors are normalized to 1. Among all the approximate binary eigenvectors that correspond to the close-to-zero eigenvalues, some contain more 1's than others. The key step of outlier detection is to eliminate those data points that correspond to the non-zero elements of binary eigenvectors with very few 1's, because the components composed of these points are very weakly connected to other components, and thus are more likely to be outliers. 
      
    Therefore, finding outliers is approximately equivalent to finding protruding 1's in the eigenvectors with close-to-zero eigenvalues. Thus, we have reduced a problem of high-dimensional outlier detection with a complex hypothesis into a one-dimensional outlier-detection problem. However, since the eigenvectors that correspond to close-to-zero eigenvalues are not perfectly binary, more elaborate methods need to be employed to detect outliers.
	
	First, eligible eigenvectors (approximate binary eigenvectors) are selected from the collection of eigenvectors. Specifically, we take the eigenvectors that correspond to eigenvalues less than $0.1$ as the candidates. Then, keeping only the binary eigenvectors, the ``high frequency" eigenvectors are excluded. Note that $\mathbf{f}$ is a ``high frequency" eigenvector if $(\sum_j|\mathbf{f}_j|-|\sum_j\mathbf{f}_j|)/\sum_j|\mathbf{f}_j|$ is sufficiently large, i.e. those eigenvectors with both large positive and negative elements. After that, the one-dimensional outlier detection algorithm is employed. For each eligible eigenvector:
   \begin{enumerate}
   \item Randomly choose half of the elements, and assign them as inliers;
   \item Find the 25\% quantile $\alpha_{1/4}$, median $\mu$, and 75\% quantile $\alpha_{3/4}$ of the selected elements;
   \item Create an interval $I=[\mu - \gamma(\mu-\alpha_{1/4}), \mu + \gamma(\alpha_{3/4}-\mu)]$ (usually select $\gamma \approx 2\sim 3$), and test all the elements: if the elements fall into $I$, then reclassify them as inliers, otherwise, reclassify them as outliers; 
   \item Repeat Step 2 and Step 3 until the classification of outliers and inliers does not change;
   \item Output the selected outliers.
   \end{enumerate}
   
   The algorithm above does not guarantee the ``correct" result. It can be helpful to run it several times and choose the result with minimum intra-class deviation (as in the case of RANSAC). However, in our simulations, we ran the routine only once, which was good enough for our data.

\subsection{Refine the Results Using RANSAC}
The algorithm described in the previous two sections uses only the proximity information of the data points. However, it is incapable of detecting type-B outliers in Fig. \ref{ClassAB}, and may easily lead to errors when the outliers are close to the inliers, even if they deviate conspicuously from the ellipse model.

To eliminate type-B outliers, we need to use the additional prior knowledge that the inliers are located on an ellipse. We can achieve this goal by running an algorithm similar to RANSAC, which has been shown to be inappropriate when the percentage of outliers is high and the number of parameters in the model is large. However, since we have greatly decreased the percentage of outliers in the remaining data set by employing the outlier detection algorithm based on proximity, it is now feasible to run a RANSAC-type algorithm.

To make our algorithm more efficient, we slightly modify the vanilla version of RANSAC \cite{Fischler1981}. First, we use the \emph{entire} set of inliers selected by the first stage algorithm to fit an initial model, instead of randomly choosing the minimum number of points as in the original RANSAC, since the remaining outliers represent just a small percentage and are close to the inliers. Moreover, since our initialization is not random, it is unnecessary to run RANSAC repeatedly. We summarize the revised model-based outlier detection algorithm as follows:
\begin{enumerate}
\item Fit a model $h$ to the ``inliers" selected by the proximity-based outlier detection algorithm;
\item Test all the data points with respect to $h$, classify the points that saliently deviate from $h$ (above a threshold) as outliers and then classify other points as inliers;
\item Refit model $h$ based on the updated inliers;
\item Repeat step 2 and step 3 until the classification does not change any further.
\end{enumerate}

With this model-based outlier detection algorithm, we are able to detect most of the missed outliers in the first stage (usually harder ones very close to the inliers), and also clear the labels for misclassified inliers.

\section{Simulation Results}
\subsection{A Typical Ellipse Fitting with Outlier Elimination}
\begin{figure}[!hbtp] 
        \centering{
        \includegraphics[scale = 0.25]{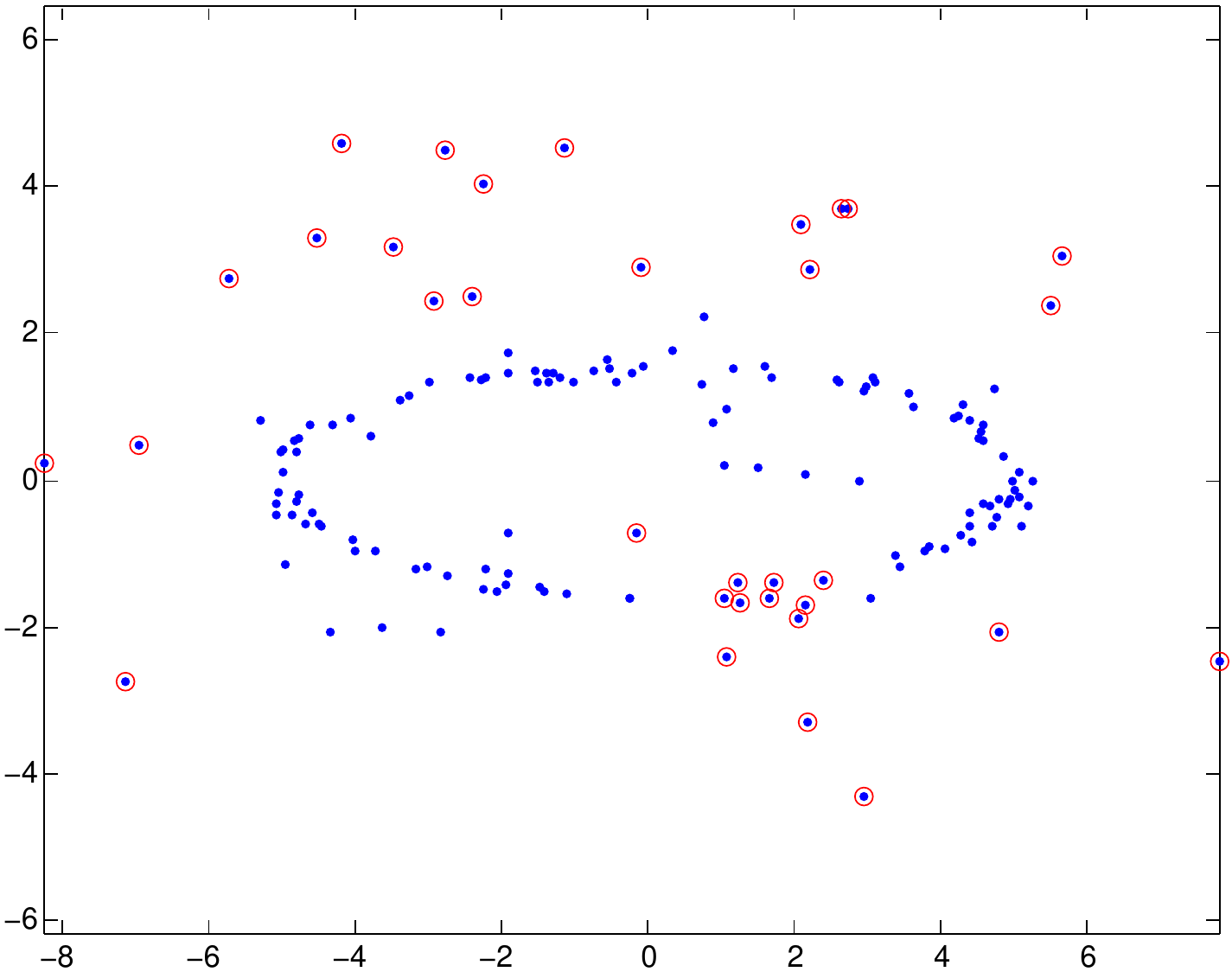}~~~~~~
        \includegraphics[scale = 0.238]{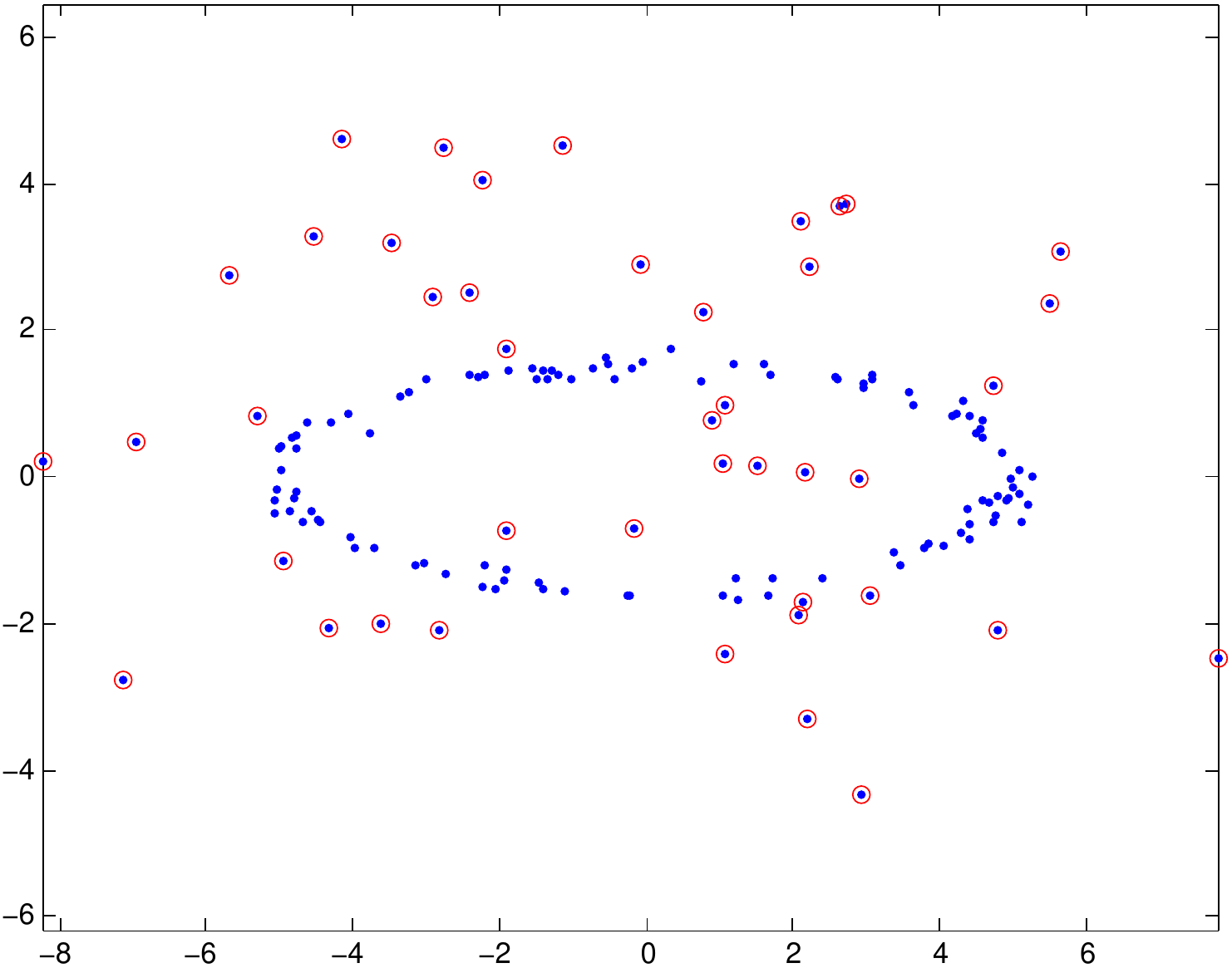}
        }
        \caption{Outlier detection results for ellipse fitting. The circled data points are classified as outliers by the algorithm. The figure on the left demonstrates the detected outliers after the proximity-based algorithm, where there are several missed outliers close to the ellipse, while a few inliers are misclassified as outliers. The figure on the right shows that these mistakes are corrected by the model-based algorithm.}
        \label{Wild2}
    \end{figure}

It is of interest to inspect a typical simulation for outlier detection of an ordinary ellipse to see the performance of our two-stage algorithm. In the simulation, we have  $N = 100$ inliers and $M = 50$ outliers. The variance of the independent Gaussian additive noise of inliers in both $x$ and $y$ directions is $\sigma_0 = 0.01$, and we simply model the outliers as data points added by an unusually large Gaussian noise, with $\sigma_1 = 2$. The true ellipse has an eccentricity $\epsilon = 0.95$ with semi-major length $a = 5$, and takes the standard position (centered at the origin, with semi-major axis aligned with the $x$ axis). The outlier detection results are shown in Fig. \ref{Wild2}.	

It is interesting to notice that the first stage of our algorithm makes several mistakes, with several outliers missed (the ones inside and closely outside the ellipse) and a few inliers misclassified (several circled points on the perimeter of the ellipse). These mistakes are corrected by the second-stage model-based algorithm. And the final result leaves us a group of purged, low-noise points from an ellipse.

\subsection{Improvement and Comparison}
\begin{figure}[!hbtp]
        \centering{
        \includegraphics[scale = 0.19]{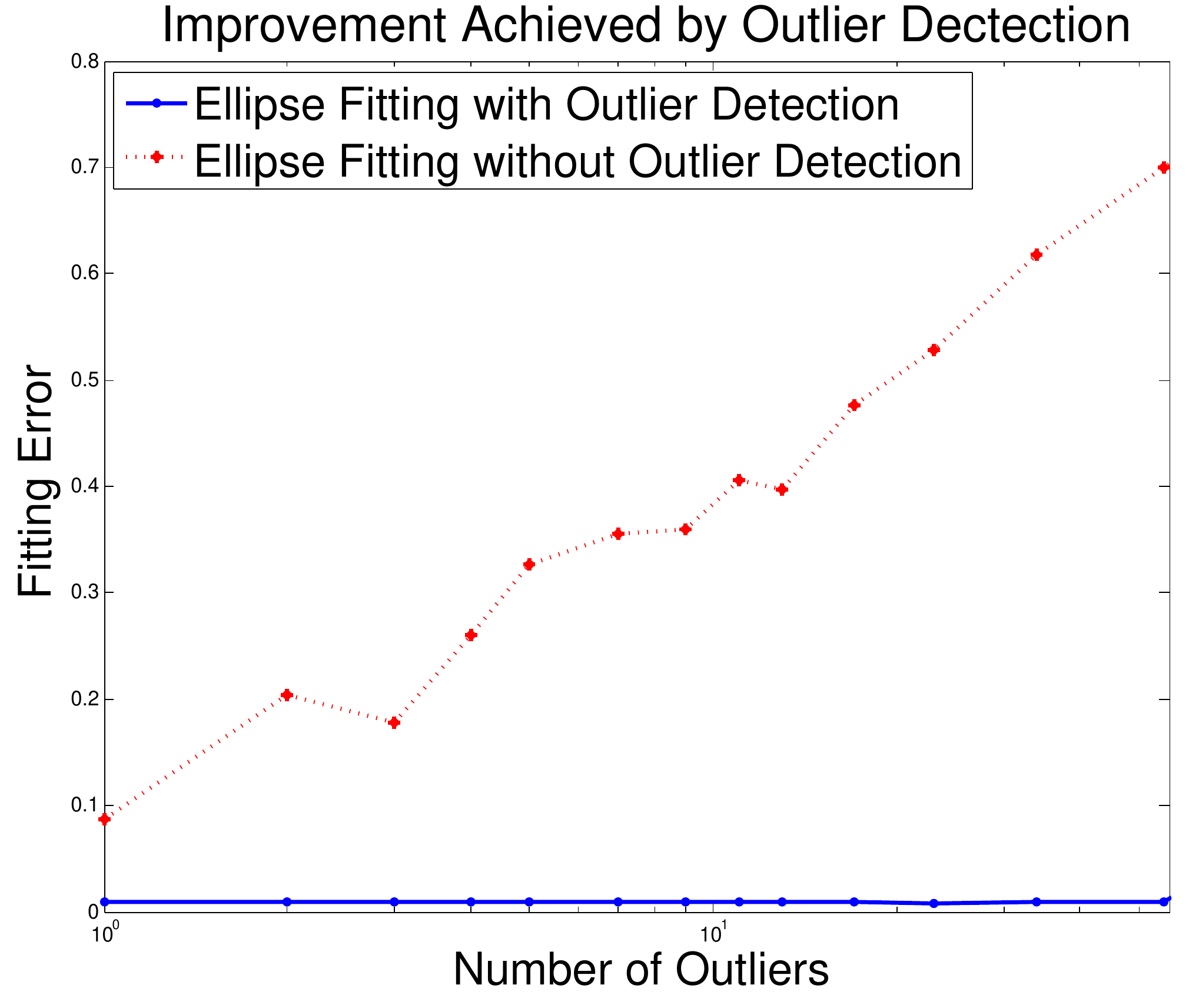}
        \includegraphics[scale = 0.19]{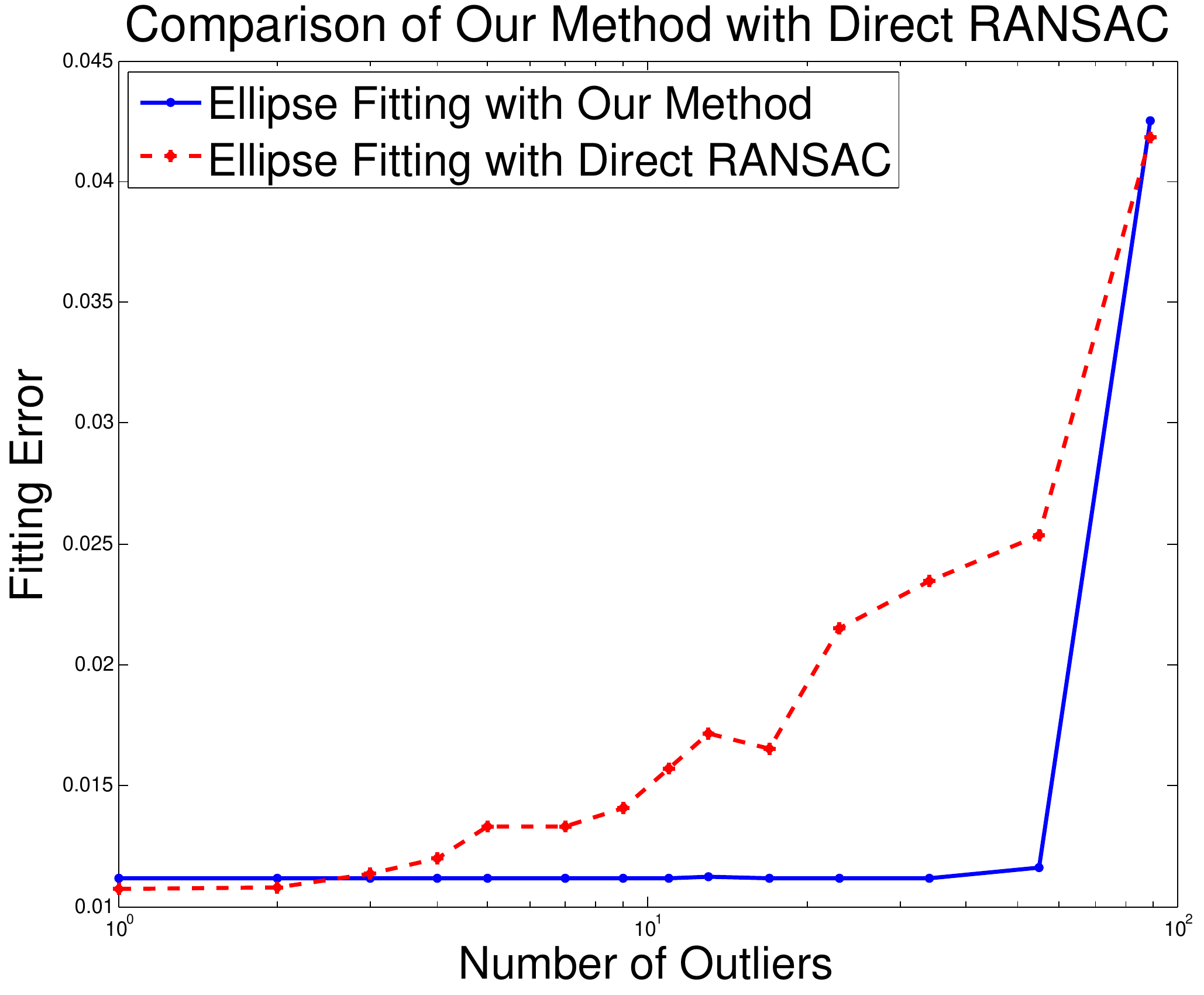}}
        \caption{Left: fitting errors with and without outlier elimination. Right: comparison between the fitting error of our algorithm and that of RANSAC.}
        \label{Percentage}
    \end{figure}

It is worthwhile to compare the performance of fitting procedures with and without outlier elimination for different numbers of outliers. Here, we choose $N = 100$, $M$ from $1$ to $55$, $\sigma_0 = 0.1$, and $\sigma_1 = 3$. The true model for the ellipse remains the same. The comparison results are shown in Fig. \ref{Percentage} on the left.
We measure the errors using the non-overlapped area of the fit and true ellipses (normalized by the area of the true ellipse) in the simulation. And the curves demonstrate the average performance of the two approaches. It is worth pointing out that for most of the cases, the performance of the fitting algorithm with outlier detection does not deteriorate as the number of outliers increases, indicating an almost perfect detection rate of the outliers. On the other hand, without outlier detection, the fitting results are not acceptable even when the number of outliers is less than 10.

We also compared our algorithm with simple RANSAC \cite{Fischler1981}, as shown in Fig. \ref{Percentage} on the right. We chose $\sigma_1 = 5$, and varied the number of outliers $M$ from $1$ to $90$. Our algorithm performs consistently well until the percentage of outliers reaches $50\%$; on the contrary, the performance of RANSAC algorithm with $1000$ iterations, though good for the cases with a small number of outliers, deteriorates rapidly as the total number of outliers increases. Our algorithm demonstrates more robustness in the presence of large numbers of outliers.

There is another robust algorithm, Five Point Fit Ellipse Fitting \cite{Rosin1993}, where all five-tuples of points are selected and fit by ellipses, the median of the parameters of which being the final fitting result. In terms of robustness, this algorithm performs competitively with our algorithm; however, it turns out to be so much more intensive in computation that it can hardly be implemented in a reasonable period of time.

\subsection{Robustness with Different Types of Outliers}
\begin{figure}[!hbtp] 
        \centering{
        \includegraphics[scale = 0.23]{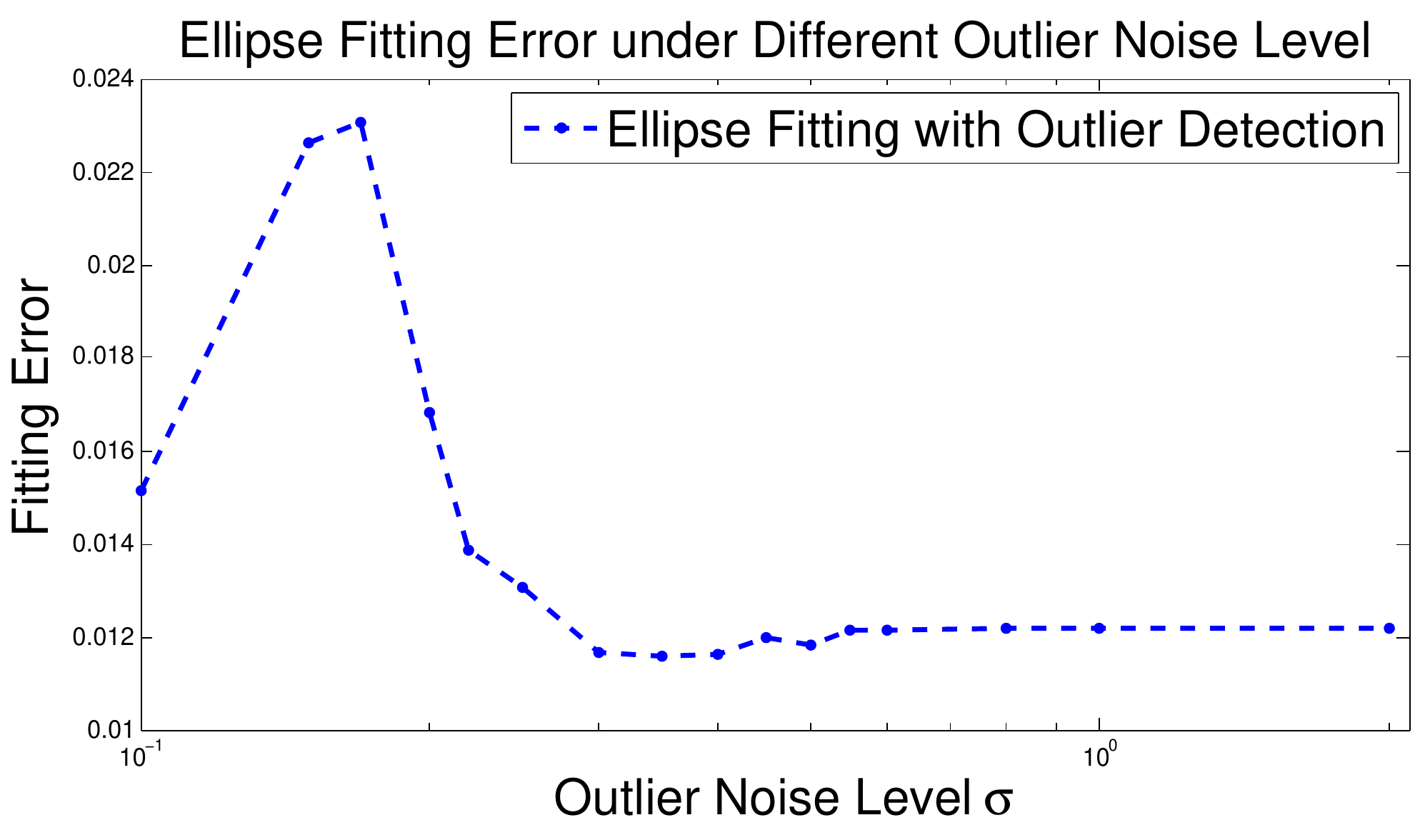}
        }
        \caption{The performance of our two-stage algorithm under a variety of outlier noise levels. Our algorithm performs consistently well for outliers with very different magnitudes.}
        \label{NoiseLevel}
    \end{figure}

Because our outlier detection algorithm is a hybrid, it is able to tackle a broad variety of outliers and the fitting error can be bounded. Here we run our algorithm for different types of outliers by adjusting the noise level of outliers $\sigma_1 \in [0.1,1.2]$, with $N = 120$, $M = 90$, $\sigma_0 = 0.1$, and the true model for the ellipse unchanged. The interesting result is shown in Fig. \ref{NoiseLevel}, where for outliers closer to the inliers as well as ouliers distant from the inliers, our algorithm performs consistently well, with fitting error tightly bounded below a low level. This shows the strong robustness of our scheme.

\subsection{Generalization to 3-D: Ellipsoid Fitting}
\begin{figure}[!hbtp] 
        \centering{
        \includegraphics[scale = 0.25]{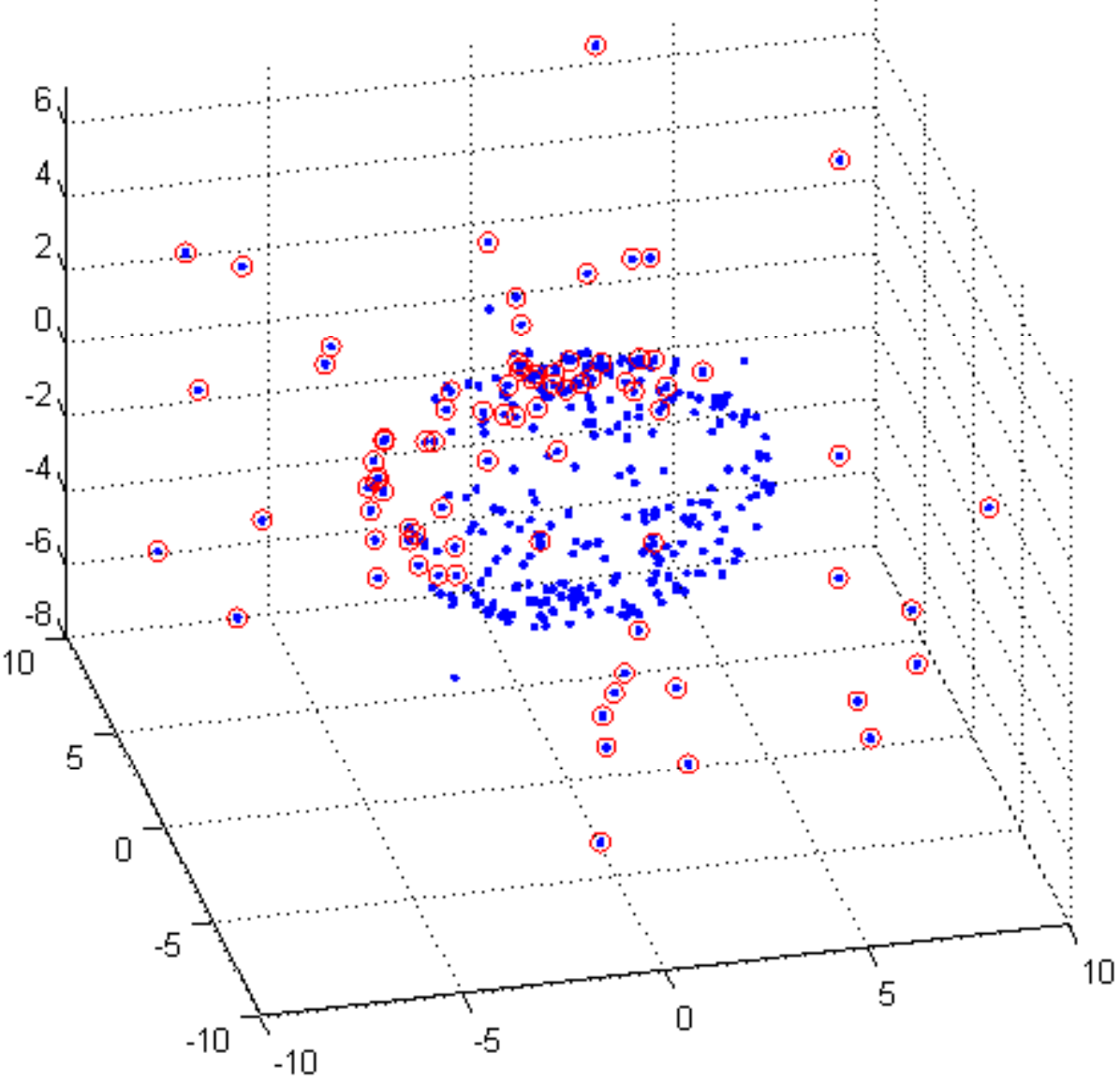}~~~~~~
        \includegraphics[scale = 0.25]{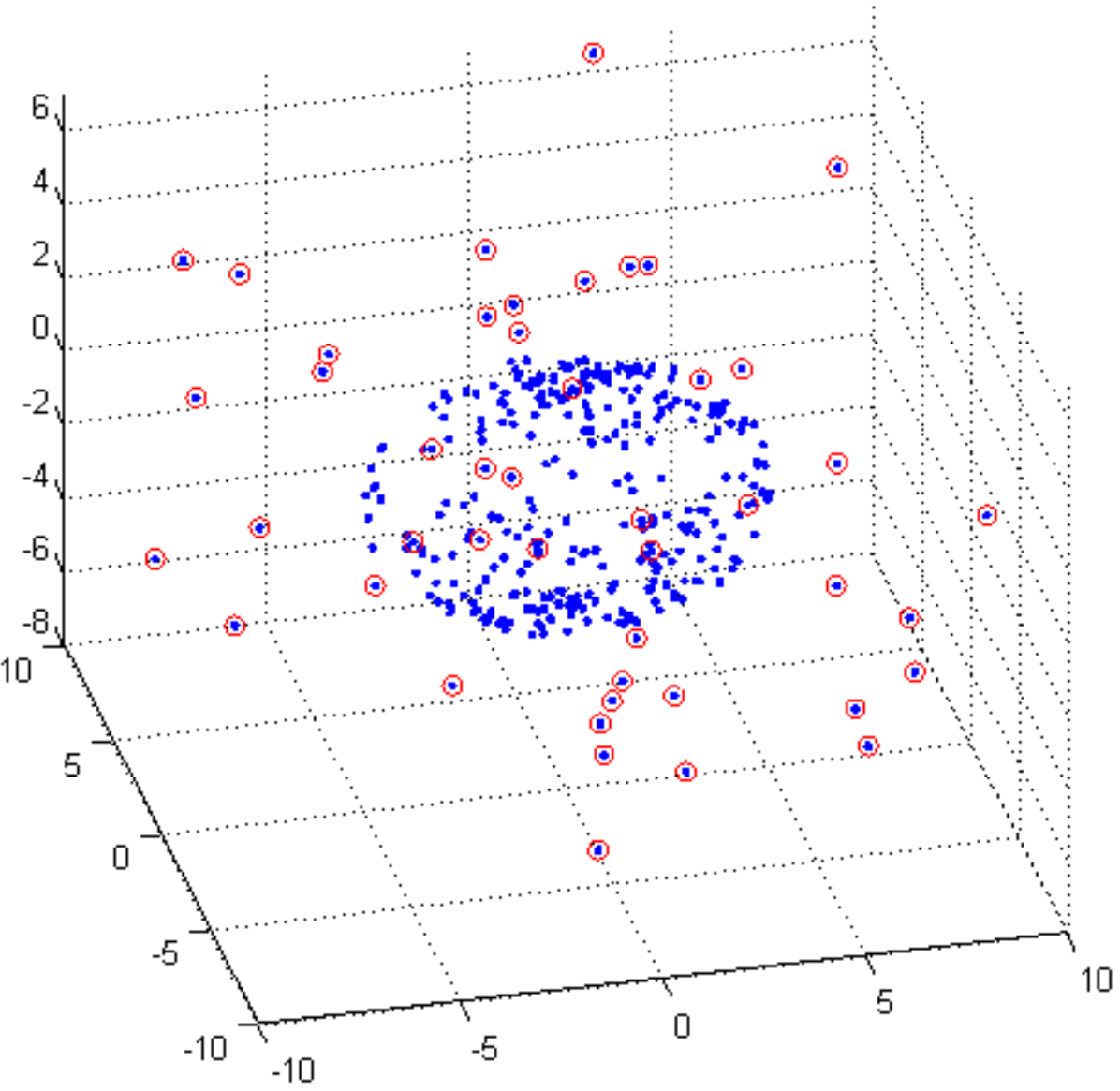}
        }
        \caption{Outlier detection results for ellipsoid fitting. Similarly to the ellipse fitting case, the proximity-based algorithm eliminates most of the distant outliers, yet misclassifies some inliers as outliers, with several close-by outliers missed, as illustrated on the left figure. These errors are corrected by the second stage model-based algorithm, as shown in the right figure.}
        \label{3D}
    \end{figure}

The basic set up for the ellipsoid fitting simulation is as follows: for inliers, $N=300$ and $\sigma_0 = 0.1$; for outliers, $M = 50$ and $\sigma_1 = 5$; the ellipsoid takes the standard position, with semi-axis lengths $a = 5$, $b = 4$, and $c = 3$. The outlier detection results are as shown in Fig. \ref{3D}.

\section{Conclusion}
Proximity-based outlier detection algorithms are good for cases in which the outliers are wildly contaminated and large in number. However, such algorithms work poorly for outliers that are close to the inliers, even though these outliers are obviously not consistent with the model; on the contrary, model-based algorithms are very good at detecting a small portion of outliers that are not consistent with the model, yet if the percentage of outliers is high and the number of parameters for the model is large, the implementation of model-based algorithms can be costly. In the problem of ellipse/ellipsoid fitting with many outliers, by combining these two types of algorithms, we have found a promising method that performs robustly with high accuracy for a variety types and numbers of outliers. We expect that this method can be generalized to other similar fitting problems, with large numbers of outliers and models with many parameters.

\bibliographystyle{IEEEtran}
\bibliography{CAMSAP}

\begin{thebibliography}{1}
\providecommand{\url}[1]{#1}
\csname url@samestyle\endcsname
\providecommand{\newblock}{\relax}
\providecommand{\bibinfo}[2]{#2}
\providecommand{\BIBentrySTDinterwordspacing}{\spaceskip=0pt\relax}
\providecommand{\BIBentryALTinterwordstretchfactor}{4}
\providecommand{\BIBentryALTinterwordspacing}{\spaceskip=\fontdimen2\font plus
\BIBentryALTinterwordstretchfactor\fontdimen3\font minus
  \fontdimen4\font\relax}
\providecommand{\BIBforeignlanguage}[2]{{%
\expandafter\ifx\csname l@#1\endcsname\relax
\typeout{** WARNING: IEEEtran.bst: No hyphenation pattern has been}%
\typeout{** loaded for the language `#1'. Using the pattern for}%
\typeout{** the default language instead.}%
\else
\language=\csname l@#1\endcsname
\fi
#2}}
\providecommand{\BIBdecl}{\relax}
\BIBdecl

\bibitem{Fitzgibbon1999}
A.~Fitzgibbon, M.~Pilu, and R.~B. Fisher, ``Direct least square fitting of
  ellipses,'' \emph{IEEE Transactions on Pattern Analysis and Machine
  Intelligence}, vol.~21, no.~5, pp. 476--480, 1999.

\bibitem{Ahn2001}
S.~J. Ahn, W.~Rauh, and H.~J. Warnecke, ``Least-squares orthogonal distances
  fitting of circle, sphere, ellipse, hyperbola, and parabola,'' \emph{Pattern
  Recognition}, vol.~34, no.~12, pp. 2283--2303, 2001.

\bibitem{Chojnacki2000}
W.~Chojnacki, M.~J. Brooks, A.~van~den Hengel, and D.~Gawley, ``On the fitting
  of surfaces to data with covariances,'' \emph{IEEE Transactions on Pattern
  Analysis and Machine Intelligence}, vol.~22, no.~11, pp. 1294--1303, 2000.

\bibitem{Hodge2004}
V.~J. Hodge and J.~Austin, ``\BIBforeignlanguage{English}{A survey of outlier
  detection methodologies},'' \emph{\BIBforeignlanguage{English}{Artificial
  Intelligence Review}}, vol.~22, no.~2, pp. 85--126, 2004.

\bibitem{Ramaswamy2000}
S.~Ramaswamy, R.~Rastogi, and K.~Shim, ``\BIBforeignlanguage{English}{Efficient
  algorithms for mining outliers from large data sets},''
  \emph{\BIBforeignlanguage{English}{SIGMOD Record (ACM Special Interest Group
  on Management of Data)}}, vol.~29, no.~2, pp. 427--438, 2000.

\bibitem{Fischler1981}
M.~A. Fischler and R.~C. Bolles, ``Random sample consensus: A paradigm for
  model fitting with applications to image analysis and automated
  cartography.'' \emph{Communications of the ACM}, vol.~24, no.~6, pp.
  381--395, 1981.

\bibitem{Mohar1991}
B.~Mohar, ``The {L}aplacian spectrum of graphs,'' in \emph{Graph Theory,
  Combinatorics, and Applications}, vol.~2.\hskip 1em plus 0.5em minus
  0.4em\relax Wiley, New York, 1991, pp. 871--898.

\bibitem{Rosin1993}
P.~Rosin, ``\BIBforeignlanguage{English}{Ellipse fitting by accumulating
  five-point fits},'' \emph{\BIBforeignlanguage{English}{Pattern Recognition
  Letters}}, vol.~14, no.~8, pp. 661--669, 1993.

\end{thebibliography}

\end{document}